\documentclass[11pt,twoside]{article}
\usepackage{asp2010}
\usepackage{natbib}

\resetcounters
\aspvolume{456}
\aspcpryear{2012}
\aspvoltitle{Fifth Hinode Science Meeting: Exploring the Active Sun}
\aspvolauthor{L. Golub, I. D. Moortel, and T. Shimizu, eds.}

\begin{document}

\title{Spectroscopic Diagnosis of Propagating disturbances in coronal loops:
Waves or flows?  }
\author{Tongjiang Wang$^{1,2}$, Leon Ofman$^{1,2,3}$, and Joseph M. Davila$^2$
\affil{$^1$Physics Dept., Catholic University of America, Washington, DC 20064, USA}
\affil{$^2$NASA Goddard Space Flight Center, Code 671, Greenbelt, MD 20771, USA}
\affil{$^3$Visiting Associate Professor, Tel Aviv University, Israel}
}

\begin{abstract}
The analysis of multiwavelength properties of propagating disturbances (PDs) using Hinode/EIS 
observations is presented. Quasi-periodic PDs were mostly interpreted as 
slow magnetoacoustic waves in early studies, but recently suggested to be intermittent 
upflows of the order of 50$-$150 km~s$^{-1}$ based on the Red-Blue (RB) asymmetry analysis of
spectral line profiles. Using the forward models, velocities of the secondary component 
derived from the RB analysis are found significantly overestimated
due to the saturation effect when its offset velocities are smaller than the Gaussian width.
We developed a different method to examine spectral features of the PDs. This method is assuming 
that the excessive emission of the PD profile against the background (taken as that prior to the PD) 
is caused by a hypothetic upflow. The derived LOS velocities of the flow are on the order 
of 10$-$30 km~s$^{-1}$ from the warm (1$-$1.5 MK) coronal lines, much smaller than those inferred from the RB
analysis. This result does not support the flow interpretation but favors of the early
wave interpretation.
\end{abstract}

\section{Introduction}
Propagating bright disturbances (PDs) in coronal loops were first observed by SOHO/EIT \citep{ber99}, 
and then by TRACE \citep{dem00}. Now SDO/AIA shows this phenomenon everywhere on the Sun from active regions 
to quiet Sun, and coronal holes \citep{tan11a, tan11b}. Whether they are flows or slow magnetoacoustic waves 
is under strong debate. The right interpretation is very important for 
our understanding of the coronal heating and acceleration of the solar wind. The wave interpretation has  
dominated for a long time until the recent discovery of persistent upflows near the loop footpoints 
by Hinode/EIS \citep[e.g][]{war11}. Now increasing studies tend to agree
with the interpretation by intermittent upflows \citep{mci09, tan11c}. 
The main evidence is based on the association of PDs with the underlying upflows, and the other 
evidence is the correlation of PDs with the blueward asymmetry and line width broadening of EIS spectra. 
However, this puzzle is still far from being clearly understood for several reasons.

\begin{figure}[!ht]
\plotone{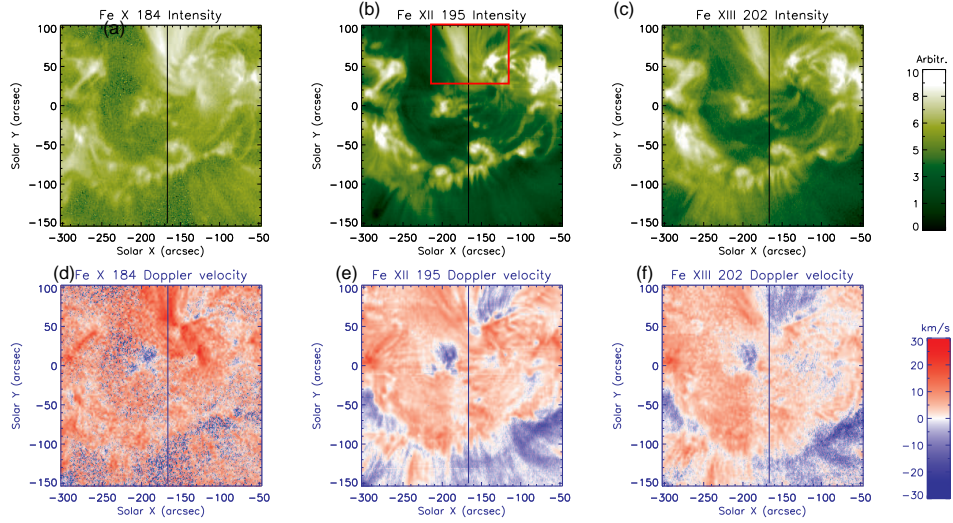}
\caption{EIS observations of fanlike coronal loops (marked with a red box in (b))
in AR 10940 taken at 00:12 UT on 2007 February 1 in three emission lines,
(a) Fe\,{\sc{x}} 184.53 \AA, (b) Fe\,{\sc{xii}} 195.12 \AA, and (c) Fe\,{\sc{xiii}} 202.05 \AA.
(d)-(f) The corresponding Doppler velocity maps derived from single Gaussian fits.
 \label{fgimg}}
\end{figure}

Observations have shown that the phenomena of upflows and PDs are closely related, 
but also appear different. 1) Upflows are continuous, whereas PDs are quasi-periodic \citep{nis11}.
2) Upflows have the higher speed at the loop base, whereas PDs can reach high altitudes 
at a nearly constant speed. 3) Upflows are progressively larger at higher temperature
up to 2 MK \citep{del08}, whereas PDs are the most obvious in typical coronal temperature (1$-$1.5 MK)
lines. 4) \citet{wan09} interpreted the PDs as slow magnetoacoustic waves, 
because the observed Doppler velocity and intensity variations have small amplitudes and 
show an inphase relationship, whereas \citet{dep10} suggested that they can be explained by weak, 
quasi-periodic high-speed upflows on the order of 50$-$150 km~s$^{-1}$. The motivation of 
this study is to examine this flow scenario using both the RB asymmetry analysis and a new proposed method.

\section{Observations}
The EIS observations were obtained on 2007 February 1 in AR 10940. \citet{wan09} first analyzed the propagating 
disturbances in Fe\,{\sc{xii}} 195.12 \AA, and found the 12 and 25 min periodicities. \citet{dep10} analyzed
the PDs in Fe\,{\sc{xiii}} 202.04 \AA\ and Fe\,{\sc{xiv}} 274.20 \AA\ using the `Red Minus Blue' line asymmetry 
technique. Figure~\ref{fgimg} shows the raster images of Fe\,{\sc{x}} 184.53 \AA, Fe\,{\sc{xii}} 195.12 \AA, 
and Fe\,{\sc{xiii}} 202.04 \AA, and their Doppler velocity maps. The upward PDs are observed in a bundle of
fanlike coronal loops. The bright fanlike loops show redshifts (downflow) in Fe\,{\sc{x}},
while the bright part shows redshifts and the weak part shows blueshifts in Fe\,{\sc{xii}} 
and Fe\,{\sc{xiii}}.  

\begin{figure}[!ht]
\plotone{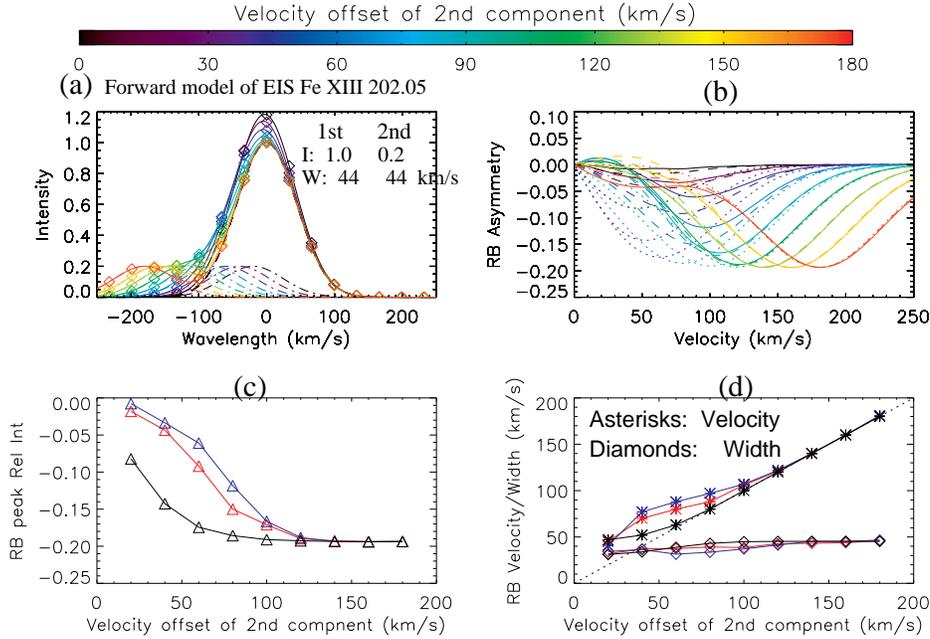}
\caption{Parameters derived using the RB asymmetry analysis for the modeled Fe\,{\sc{xiii}} 
202.05 \AA\ line. (a) The primary component (dashed line), secondary component (dot-dashed line),
and their total (solid line) emission profiles. The Gaussian width (W) and peak intensity (I) of 
the primary and secondary components are shown on the plot. 
(b) RB$_S$ (solid), RB$_P$ (dashed), and RB$_D$ (dotted) asymmetry profiles (see the text). 
(c) Variations of the negative peak intensity of the RB asymmetry profile for different 
offset velocities.  (d) Variations of the RB velocity ({\it Asterisk}) and RB Gaussian width 
({\it diamond}) of the secondary component. The blue, red and black curves
represent the results from RB$_S$, RB$_P$, and RB$_D$, respectively.
 \label{fgmod}}
\end{figure}

\section{The Saturation Effect of the RB technique}
The RB technique was developed by \citet{dep09} to quantify the line asymmetry. The RB asymmetry is defined as the emission difference between the red and blue wings integrated over the same wavelength range. By calculating 
the RB asymmetry at different velocity offsets, a RB profile is obtained, from which parameters of the asymmetry such as velocity offset, line width, and relative amplitude can be derived. \citet{mar11}
quantitatively studied the properties of the RB asymmetry using both simple forward models and 3-D
MHD simulations. 

\begin{figure}[!ht]
\plotone{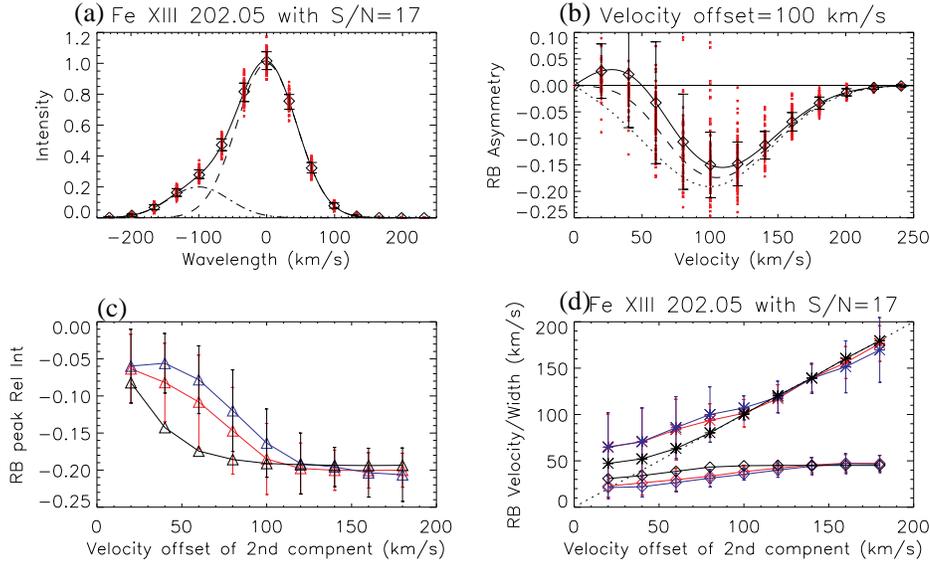}
\caption{Forward modeling of (a) the EIS Fe\,{\sc{xiii}} 202.05 \AA\ line profile and (b)
the RB asymmetry profile. The data points and error bars are the mean values and
standard deviations for 100 different realizations by adding photon noises with S/N=17.
In (b) the solid, dashed and dotted curves are for RB$_S$, RB$_P$, and RB$_D$.
(c) and (d) Same as those in Fig.~\ref{fgmod} but for the case with photon noises.
 \label{fgfwd}}
\end{figure}

Figure~\ref{fgmod} shows the forward models of the RB asymmetry profiles calculated using three 
ways (called RB$_S$, RB$_P$, and RB$_D$), which are similar to those used in \citet{tan11b}.
RB$_S$ is calculated with the line centroid derived from the single Gaussian fit to the line core,
RB$_P$ with the line centroid taken as the spectral peak position, and RB$_D$ with the line centroid 
derived from the double Gaussian fit (in the models taken as the exact centroid position of 
the primary component). It shows that the derived RB velocity tends saturated for the small flow velocity 
(Fig.~\ref{fgmod}d). This effect results from two reasons. 1) The spectral line has a certain width. 
When the velocity offset is close to the Gaussian width, a non-negligible amount of emission from 
the secondary component contributes to the red wing of the primary component, thus distorts 
the RB profile. 2) When the velocity offset is small, the determination of the line centroid for 
the primary component becomes unreliable for any methods. The derived line centroid position is 
largely deviated from the true one. The saturation effect leads to the flow amplitude underestimated 
and the flow velocity overestimated significantly.

\begin{figure}[!ht]
\plotone{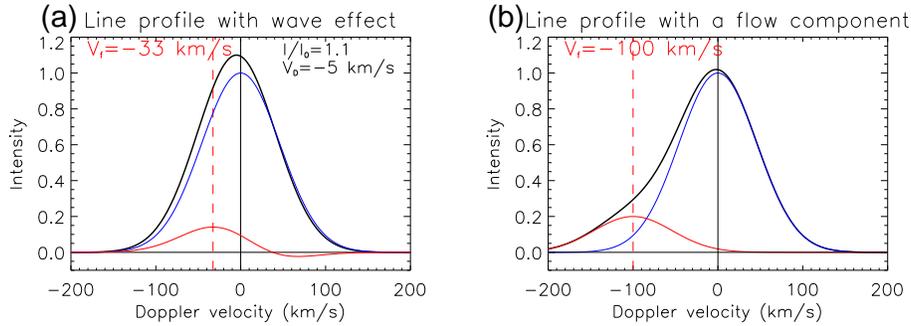}
\caption{Forward modeling of the line profiles of a bright propagating disturbance caused by 
(a) a slow-mode wave, and (b) a high-speed upflow.
 \label{fgpfm}}
\end{figure}

\begin{figure}[!ht]
\plotone{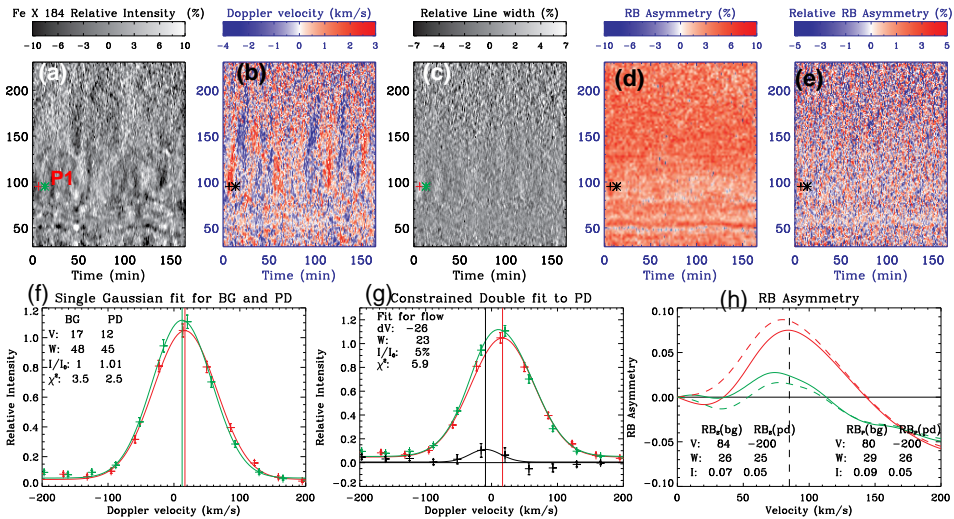}
\caption{ Spectral analysis of PDs observed in Fe\,{\sc{x}} 184.53 \AA.
(a)-(e) Time series of relative intensity, relative Doppler shift, relative Gaussian width, 
60--110 km~s$^{-1}$ RB asymmetry, and relative RB asymmetry.
(f) Single Gaussian fits to the line profiles for the PD (green) and the background (red) 
at position P1 marked in (a).  (g) Gaussian fit to the remnants (black)
of the PD profile exceeding the background profile, which is assumed to be caused
by a {\it hypothetical} upflow.  (h) RB asymmetries for the PD profile
(green) and the background profile (red). The solid lines and the dashed lines represent 
RB$_{S}$ and RB$_{P}$, respectively. The measured parameters for the RB asymmetry peak 
are shown on the plot.
 \label{fgf10}}
\end{figure}

The saturation effect becomes more serious for data with the noise. Figure~\ref{fgfwd} shows the forward 
models including Poisson noises based on the S/N of EIS data. The derived RB velocities 
have large uncertainties for small velocity offsets especially in the case with the small S/N. 
The simulations for 6 coronal lines show that for an upflow with the velocity offset of 20 km~s$^{-1}$, 
the RB velocities from RB$_S$ and RB$_P$ are 55$-$75 km~s$^{-1}$. The consequences of the saturation 
effect are, 1) Due to the large instrumental line broadening, the RB velocity derived from EIS data 
will be never below 50 km~s$^{-1}$. 2) The so-called RB-guided double Gaussian fit will be misguided to 
overestimate the upflow velocity when it is actually small, leading to a biased conclusion that 
upflow velocities are always on the order of 50$-$150 km~s$^{-1}$, independent on the temperature.

\section{Spectral Signature of the Propagating Disturbances}
We suggest a more simple method to examine whether the bright PDs are intermittent flows. If they are, 
we take the line profile from the time between two PDs as the background, and subtract it 
from the PD line profile, then the obtained remnant emission should come from the transient flow.  
The advantage of this method is whatever the background profile is (symmetrical or asymmetrical),
as long as it is slowly-varying, after being removed the left emission should be
the transient component that makes the PDs. Figure~\ref{fgpfm} compares the modeled line profiles
between the wave and flow cases. In the wave case, the PD profile is modeled by blueshifting the
background profile by 5 km~s$^{-1}$ and increasing the peak intensity by 10\%. The obtained remnant 
profile has a peak at 33 km~s$^{-1}$ in the blue wing, and shows weak negative intensities in
the red wing since the wave causes the line core blueshifted in the compressive phase. 
In the flow case the PD line profile
is modeled by adding a high-speed (100 km~s$^{-1}$) upflow component, showing a clear blueward asymmetry.
We expect that the remnant line profile should not have negative values in the flow case.

Figure~\ref{fgf10} shows the analysis of the PDs in Fe\,{\sc{x}} 184 \AA. The PDs in intensity 
are weak due to the small S/N, but are obvious in Doppler shifts, which are similar 
to those seen in Fe\,{\sc{xii}} 195.12 \AA\ \citep{wan09}, showing the upward propagating features.
This is inconsistent with the downflows as indicated by Doppler redshifts and redward asymmetries
of Fe\,{\sc{x}} line profiles if the PDs are upflows. Instead, the remnant profile shows 
the peak with a blueshift of 26 km~s$^{-1}$ and the negative values in the red wing 
(Fig.~\ref{fgf10}g), which are consistent with the expected features from the wave model 
(Fig.~\ref{fgpfm}a). The analysis of the remnant profiles for the PDs in
Si\,{\sc{x}} 258, Fe\,{\sc{xii}} 195, and Fe\,{\sc{xiii}} 202 \AA\ shows the similar results 
which reveal no evidence for the high-speed ($\sim$100 km/s) transient upflows. 
Therefore, our results support the early interpretation of PDs by the slow magnetoacoustic wave.

\section{Discussion and Conclusions}
Using the line-profile difference analysis, we find only small (10$-$30 km~s$^{-1}$) hypothetic upflows  
in Fe\,{\sc{x}}$-$Fe\,{\sc{xiii}} and Si\,{\sc{x}}. Considering the apparent speed of the PDs of
$\sim$100 km~s$^{-1}$ \citep{wan09}, if they are flows the inclination of magnetic fields
along which the PDs are seen would be 73$^{\circ}$-84$^{\circ}$, much larger than the typical
values measured from observations \citep{mar09} or the field extrapolations \citep{tan11b}. 
\citet{tan11c, nis11} suggested the presence of quasi-periodic upflows at the base of loops
based on the blueward asymmetry of the line profiles and its correlation with the line width broadening.
This suggests that propagating slow-mode waves at higher locations may be driven by fast upflows 
at the loop's footpoint which are generated by tiny heating events (nanoflares). Using 3-D MHD simulations, 
\citet{ofm11} showed that the injection of periodic upflows can produce persistent upflows and
excite propagating slow-mode waves in the loop simultaneously.
    
In summary, we find that the blueward asymmetries of EIS spectra are largely overestimated due to the
saturation effect using the forward models including the photon noise. This may lead to a biased 
conclusion that upflow velocities are temperature-independent. The propagating disturbances are 
the most obvious seen in typical (1$-$1.5 MK) coronal lines. The RB analysis shows no evidence that
the PD line profiles have an obvious increase in large ($\sim$100 km~s$^{-1}$) blueward asymmetries 
compared to the background profile. This result is consistent with the derived small (10$-$30 km~s$^{-1}$) 
`hypothetical' upflows from the line-profile difference analysis, thus supporting 
the slow-wave interpretation of PDs.

\acknowledgements 
LO was supported by NASA grants NNX08AV88G, NNX09AG10G, and NNX10AN10G. 
TW was supported by NASA grants NNX08AE44G and NNX10AN10G.

\bibliography{wang_arXiv}

\end{document}